# High-extinction VIPA-based Brillouin spectroscopy of turbid biological media

Antonio Fiore[1], Jitao Zhang[1], Peng Shao[2], Seok Hyun Yun[2] and Giuliano Scarcelli[1,a]

[1] *Fishell Department of Bioengineering, University of Maryland, College Park, College Park, MD 20742, USA*

[2] *Harvard Medical School and Wellman Center for Photomedicine, Massachusetts General Hospital, 50 Blossom St., Boston, MA 02114, USA*

Brillouin microscopy has recently emerged as powerful technique to characterize the mechanical properties of biological tissue, cell and biomaterials. However, the potential of Brillouin microscopy is currently limited to transparent samples, because Brillouin spectrometers do not have sufficient spectral extinction to reject the predominant non-Brillouin scattered light of turbid media. To overcome this issue, we developed a spectrometer composed of a two VIPA stages and a multi-pass Fabry-Perot interferometer. The Fabry-Perot etalon acts as an ultra-narrow band-pass filter for Brillouin light with high spectral extinction and low loss. We report background-free Brillouin spectra from Intralipid solutions and up to 100 microns deep within chicken muscle tissue.

[a] *Corresponding author: scarc@umd.edu*

Brillouin light scattering spectroscopy has been a powerful technique in applied physics and material science for several decades by enabling the noninvasive characterization of material properties through the measurement of acoustic phonons[1]. From a measurement standpoint, Brillouin scattering spectroscopy is challenging because it requires both high spectral resolution to resolve optical frequency shifts on the order of 1-10 GHz (i.e. <0.001 nm) and high spectral extinction to detect weak spontaneous Brillouin signatures next to the much stronger (> $10^7$) non-shifted optical signals. To address these challenges, historically, Brillouin spectroscopy has relied on multi-pass scanning Fabry-Perot (FP) interferometers[2]. Although these instruments have been continuously refined over the years[3], the scanning-based approach and low-throughput generally result in long data acquisition times of minutes to hours per spectrum[4].

In the past few years, a new approach to Brillouin spectroscopy has emerged that dramatically enhanced measurement throughput. Using virtually imaged phased array (VIPA) etalons[5], parallel spectral detection enabled to collect the entire Brillouin spectrum in one shot with sub-GHz resolution and high throughput efficiency. This advancement has allowed performing Brillouin spectral characterizations within milliseconds and low power levels thus extending the reach of Brillouin spectroscopy to biomaterials, biological cells and ocular tissue in vivo[5,6]. However, VIPA-based spectrometers have not yet achieved the spectral extinction ratio of FP interferometers, and this has limited the interrogation of turbid media such as biological tissue. In this context, significant effort to increase spectrometer's ability to reject non-Brillouin scattered light has been put forward in the past few years[7,8].

Here, we report a spectrometer configuration featuring a tunable, high-throughput and narrow-bandpass filter based on a low-finesse Fabry Perot etalon. Thanks to this innovation, we increased the overall spectral extinction by more than 10 fold with respect to state-of-the-art spectrometers with less than ~2 dB insertion loss. This enabled us to perform rapid Brillouin spectral characterization deep into non-transparent biological tissue without any limitation due to elastic scattering background.

The spectrometer consists of a triple-pass Fabry-Perot (3PFP) bandpass filter and a two-stage VIPA spectrometer (Fig. 1a). The 3PFP filter was placed on the collimated beam path, before a two-stage VIPA spectrometer featuring two VIPA etalons of 17GHz Free Spectral Range (FSR). To build the 3PFP, we used a fused silica etalon of 3.37 mm thickness (i.e. 30GHz FSR) coated for 60% reflectivity, resulting in low finesse (6). Using low reflectivity and low finesse, the etalon in single–pass configuration has a ~5 GHz bandwidth with <10 % loss. However, at low finesse the spectral extinction is limited to 11 dB. Using a triple-pass configuration, the filter featured a 3GHz bandwidth and ~ 40% insertion loss. The throughput of the filter is less than the ideal case (<15%) because of secondary resonant cavities formed between the mirrors and the etalon due to the multi-pass configuration. Nevertheless, for equal rejection performances, the multi-pass low-finesse approach is twice as efficient as a high-finesse single pass configuration[9].

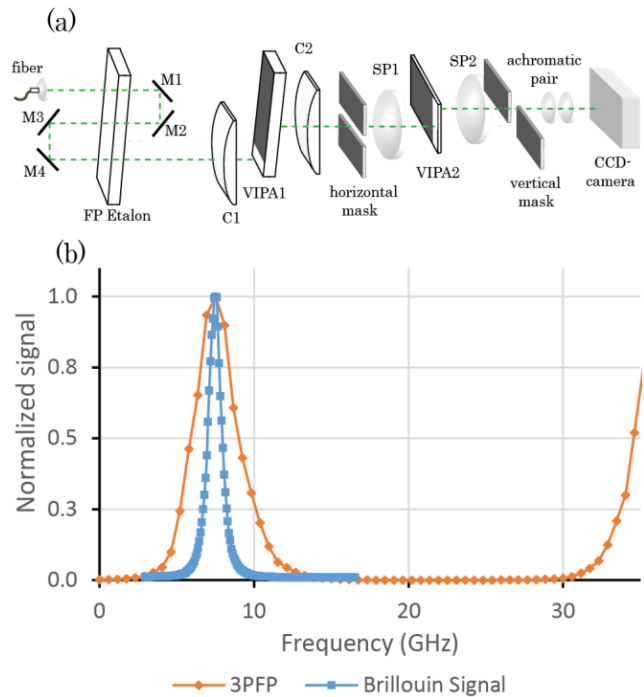

Figure 1. (a) Setup of the Brillouin spectrometer; the triple pass Fabry-Perot filter is built using two corner cubes (M1-M2 and M3-M4); (b) Comparison between the 3PFP transmission profile and the Brillouin anti-Stokes signal of a water sample. The 3PFP filter suppresses the Stokes peak.

By changing the angle between the etalon and the incoming beam, the filter bandpass can be tuned with high precision to isolate the desired Brillouin scattering signal (Stokes peak from a water sample in this case) as shown in Figure 1b. This allows to suppress the stray laser light due to reflections within the spectrometer as well as the background noise due to elastic scattering within the sample. Note that, in Figure 1b, also the Brillouin Stokes component is suppressed. However, because typical variations within samples in Brillouin measurements are less than 2

GHz, the bandwidth of the filter is large enough to perform Brillouin imaging in one setting without necessity of tuning the angle.

The spectral extinction of the spectrometer is characterized in Figure 2. To construct the curve, we performed several measurements of the two-stage VIPA spectrometer extinction while varying the angle (and thus the central frequency) of the Fabry-Perot filter. Because the transmitted intensity varied over several orders of magnitude, the limited dynamic range of the camera could not capture the whole spectral profile in one setting; therefore, we acquired several profiles at fixed laser incident power in different detection configurations by tuning gain ($g$), exposure time ($t$) and using a neutral density filter of throughput ($f$). The intensities ($I$) for each acquisition were then scaled according to the relation $I = (counts * p)/(g * t * f)$. The overall spectrometer has a maximum extinction of about 85dB with optical throughput of 16% (~58% from the 3PFP filter and 27 % from the two-stage VIPA spectrometer). This is about 15 dB larger than previous spectrometers with similar throughput[6]. Because the FSRs of the FP filter and of the VIPAs are not matched, the two VIPA peaks at 0 and 17 GHz do not have equal intensity, and the point of maximal extinction does not fall at half of the two VIPA peaks. If the FSRs of 3PFP and VIPA etalons were to be matched, the maximum extinction would be larger than 90 dB with the same overall throughput.

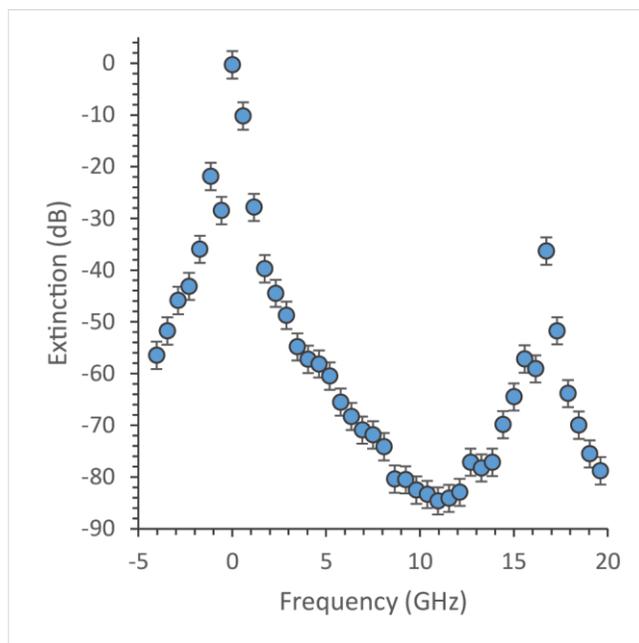

Figure 2. Extinction ratio measurement of the Brillouin Spectrometer.

To quantify the performances of our setup in the measurements of turbid media, we connected the Brillouin spectrometer to an inverted confocal microscope and measured the Brillouin spectra of intralipid solutions. Light from a 532 nm single mode laser was focused into an intralipid solution sample, placed in a transparent plastic dish. Figure 3a shows photographs of intralipid solutions of increasing concentration and increasing opacity. Intralipid solution at 10%, clearly non-transparent, are generally used to mimic the elastic scattering properties of biological tissue and thus provide a useful tool to quantify the rejection of non-Brillouin scattered light and compare it across studies. Previous works have shown that a two-stage VIPA spectrometer can suppress the elastic scattering component only up to a concentration of 0.001%; a three stage spectrometer can suppress the background light of up to a 5% Intralipid solution but with a total loss of over 90%[7].

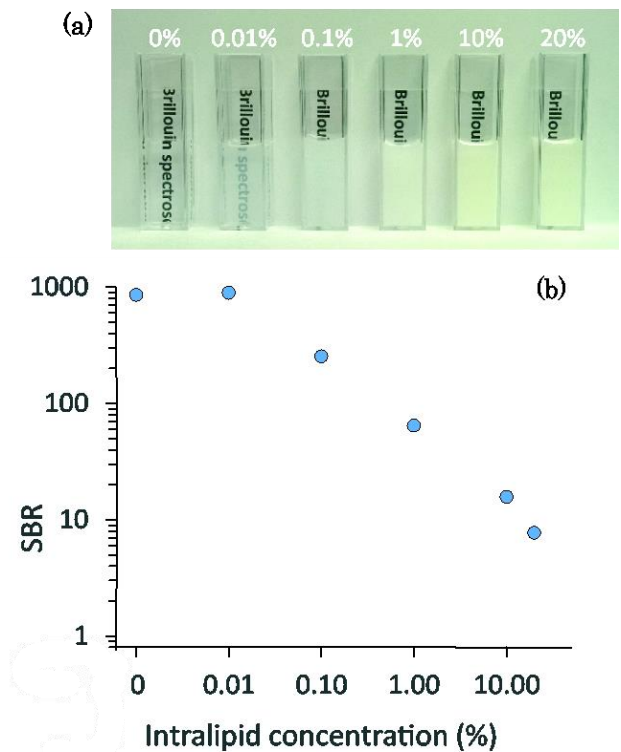

Fig 3. Measurements on intralipid solutions; the values are calculated on 100 frames average. (a) Qualitative comparison of transparency for different concentration. (b) Signal to background ratio of Brillouin spectra for different intralipid concentrations.

Figure 3b shows the signal to background ratio of a Brillouin spectrum acquired from pure water up to a 20% intralipid solution. The elastic background is very well suppressed showing an SBR greater than one at all concentrations. Importantly, we verified that at 10% intralipid solution, our Brillouin measurement is shot-noise limited, thus unaffected by elastic background.

Next, we demonstrated the ability to detect background-free Brillouin signal from biological tissue. The tissue sample was a thin slice of chicken breast of area of about 1cm$^2$, placed on the bottom of glass dish plate. The laser light was focused into the tissue through the glass cover-slip at a power of 6mW and exposure time of 300ms. Fig. 4a shows the Brillouin signal intensity as a function of tissue depth reported as average and standard deviation of 100 frames. The focusing depth was varied by translating the objective lens along the z-axis of the microscope. For comparison, the background signal at each depth is reported, showing the excellent suppression of the elastic scattering provided by the instrument. Figure 4b shows the signal to noise ratio of this measurement as a function of tissue depth. Brillouin spectra with SNR greater than one were obtained from the tissue at depths up to 100 μm. We observed an exponential decay of SNR yielding a mean free path of ~150 μm, consistent with the intensity loss due to elastic scattering in chicken breast tissue[10] and thus similar to other optical modalities that do not suffer from background issues.

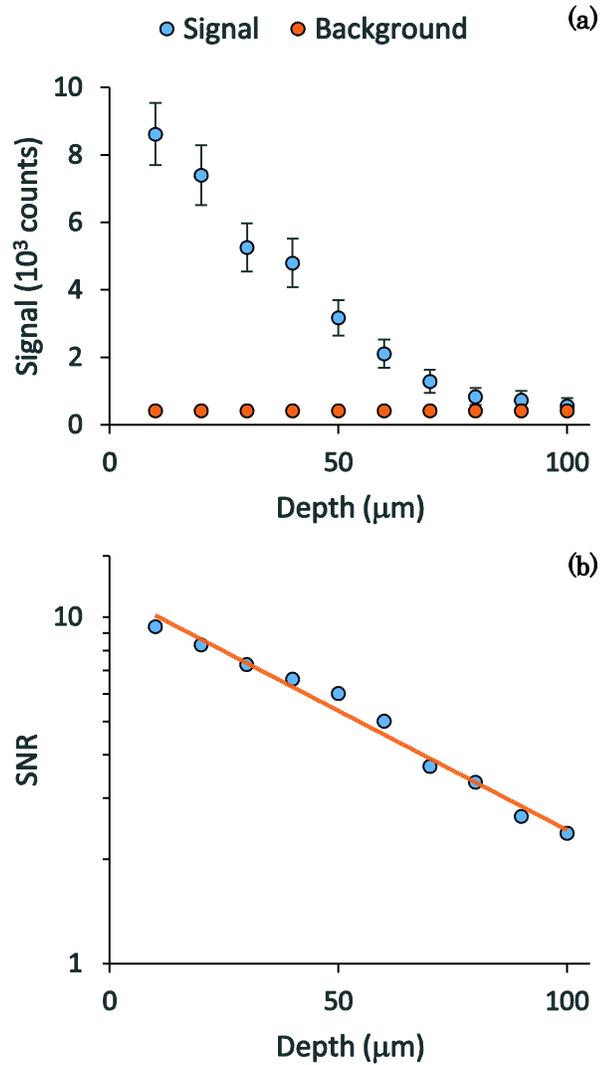

Fig 4. (a) Comparison between Brillouin signal and elastic scattering light component at different depths. (b) Signal to noise ratio as a function of depth; the SNR is calculated on 100 frames average, with an exposure time of 0.3s. The exponential fit leads to a mean free path of ~150 microns.

Importantly, other methods of rejecting elastically-scattered light that use absorption lines of gas cells require experimentalists to perform Brillouin measurements with specific laser wavelengths and are prone to degradations due to laser frequency drift. In contrast, our spectrometer maintains its performances over a spectral band only limited by the coatings of the optical elements (Fabry-Perot, VIPA, mirrors and lenses). Therefore, the spectrometer can accommodate Brillouin experiments with a wide variety of excitation wavelengths; moreover, operating at low-finesse and with a tunable etalon, laser frequency drifts can be effectively handled.

Several improvements can be implemented to improve the practical performances of our spectrometer. As mentioned before, the FSR of the Fabry-Perot etalon and the VIPAs should be the same to reach maximal extinction. Moreover, using the same principles demonstrated here, a reflection configuration of the 3PFP would allow to build a notch-filter rather than a bandpass filter, with similar performances of throughput and extinction but larger transmitted bandwidth, which could be useful for Brillouin measurements exhibiting large shift variations of non-homogenous sample.

In conclusion, we have developed a two-stage VIPA Brillouin spectrometer with confocal sampling and a triple pass Fabry-Perot bandpass filter. We demonstrated rapid Brillouin measurement of intralipid solutions and turbid biological tissue with effective suppression of elastic scattering background. This investigation may expand the reach of Brillouin technology beyond ocular tissue to important areas such as the mechanical characterizations of tumors and atherosclerotic plaques.

This work was supported in part by the National Institutes of Health (K25EB015885, R21EY023043, R01EY025454); National Science Foundation (CMMI-1537027, CBET-0853773); Human Frontier Science Program (Young Investigator Grant) and the Ministry of Science of Korea, under the "ICT Consilience Creative Program" (IITP-2015-R0346-15-1007).